# X- Ray Entangled Photon Production in Collisions of Laser Beams with Relativistic Ions

K.A. Ispirian, M.K. Ispiryan

Yerevan Physics Institute, Yerevan 0036, Armenia

A method is suggested to produce, with the help of colliding laser photons with bunches of relativistic ions having two energy levels, both intense beams of monochromatic polarized X-ray fluorescence photons and significant number of X-ray entangled photons, via double Doppler transformation. Nonlinear susceptibility of the ions, the cross section and the rate of production of such photons at RHIC are estimated. Such beams of X-ray photons can be detected and applied to solve various problems, in a manner similar to the usage of optical photons.

It is expected [1] that, after the launching of X-ray SASE FELs [2], a breakthrough will take place in the study and application of the X-ray nonlinear (NL) processes like what happened to the study of optical NL phenomena after the discovery of lasers in 1960s. At present, thousands of published works are devoted to the study of the optical NL effect of parametric down conversion (PDC) and its applications in quantum computing, cryptography, teleportation, etc. During PDC, a "pump" photon with angular frequency $\omega_P$ passes through a NL medium and decays into two photons - a "signal" photon with frequency $\omega_S$ and an "idler" photon with $\omega_i$. Intense laser beams are the source of pump photons. As NL media, NL crystals are often used, although other media, for instance, gas with relatively high value of NL susceptibility $\chi^{(2)}$ can be used as target. Such a gas, as it has been shown in a recent paper [3], has no saturation property.

PDC in X-ray region for the first time has been considered in [4] and observed in [5]. Since the cross section of PDC is small, it has been further experimentally studied only in a few works [6-9], but without satisfactory development and comparison between the theory and experiment. The highest event rate observed with the third generation synchrotron radiation sources [7-9] is around 0.1 s$^{-1}$, while the expected rate for LCLS and TESLA FELs is 10000 times higher [2]. As in the optical range, the X-ray PDC is interpreted semi-classically as mixing of the pump photons with zero point fluctuation photons, and the very small values of X-ray nonlinear susceptibilities are somewhat compensated by higher zero point fluctuation fields. X-ray PDC takes place as non-linear Bragg diffraction in crystals with energy and momentum conservation laws

$$\hbar\omega_P = \hbar\omega_S + \hbar\omega_i; \quad \vec{k}_P + \vec{H} = \vec{k}_S + \vec{k}_i, \qquad (1)$$

where $\vec{k}_j$, $j = P, S, i$ are the corresponding momenta of the photons and $\vec{H}$ is the crystal reciprocal lattice vector.

The demand for intense sources of entangled X-ray photon pairs for applications in various fields is so high that even such exotic processes as Unruh radiation [10] and inverse double Compton scattering [11] are proposed for their production. In all the works [4-11] photons from intense monochromatic X-ray beams serve as pump photons, while crystals in the Bragg geometry provide the NL susceptibility $\chi^{(2)}$. In [12] it has been proposed to use high energy electron beams for the production of X-ray entangled photons directly in crystal targets without X-ray photon beams. According to [12], part of pseudophoton cloud of the electron serves as the source of pump photons.

Taking into account the fact that the density of optical photons in laser beams is much higher than the particle densities in electron beams, as in the case of optical PDC, we propose to use laser beams colliding with relativistic ions (RI) having at least two energy levels. RIs serve as NL medium. The conversion the optical photons into X-ray photons takes place due to Doppler transformations. Indeed, as it has been shown for the first time in [13], after Doppler transformations the scattered resonance radiation photons of laser beam produce monochromatic X- and γ-beams, i.e., the RIs work as "relativistic mirrors," reflecting the laser photons with much higher cross sections than the "electron mirror" in the process of inverse Compton scattering [14,15]. In the case of head-on collisions in their rest frame (RF), the relativistic atoms or ions with energy levels $E_1$, $E_2$, $E_2 - E_1 = \hbar\omega_{21}$ "see" the laser photons with energies $\hbar\omega' = 2\gamma\omega_L$ ($\gamma = E/mc^2 = 1/\sqrt{1-\beta^2}$ is the Lorentz factor). If $\hbar\omega' = \hbar\omega_{21} = E_2 - E_1$, the laser photons can be scattered with large resonance cross section, $\sigma \sim \lambda_{21}^2$, providing fluorescence X-ray photons which in the laboratory frame (LF) have energy up to $\omega_1 = 4\gamma^2\omega_L$ and fly in the direction of RI momentum. The ideas of [13] have been developed in [16] and [17] for atomic and nuclear energy levels, respectively. In [18] it has been shown that the scattered photons have polarization approximately equal to the polarization of the laser photons.

The main reason of the fact that still no such X-ray beam production has been implemented is the absence of partly stripped RIs. Methods have been discussed [19], which could convert the fully stripped RIs of RHIC and LHC into hydrogen-like atoms with filled K-shells. Nevertheless, in [20] one can find calculation results on production of such X-ray beams for Fermilab Tevatron. Considering the particles channeled in a crystal as two- or multi-level relativistic systems, in [21] the authors have considered the resonance scattering of laser photons on such systems.

To derive further formulas, let us consider first collisions of laser photons with hydrogen-like ions with one strong resonance level with width $\Gamma$. The atoms are at rest in LF. The resonance fluorescence is dominant in such collisions if the resonance condition $E_2-E_1=\hbar\omega_{21}=\hbar\omega_L$ takes place. One can look at the process as consisting of two steps: the K (1S) electron is shifted to a vacant level near continuum after absorption of a laser photon, and then the vacant K shell is filled, emitting a fluorescence photon with very large resonance cross section [13]. The emission of an entangled photon pair due to NL susceptibility is also possible with much lower probability. The emitted photons have almost isotropic angular distribution. However, due to the NL part of the susceptibility, the PDC process can occur with emission of two photos – a signal and an idler - instead of a single fluorescence photon. For the X-ray region, one can obtain the total cross section for such a PDC process when the resonance condition is satisfied by integrating formula (11.56) of [22] over the solid angle:

$$\sigma_{res}^{Ent}(\omega_P\to\omega_S+\omega_i)=\frac{1}{72\pi^2(\hbar c)^7}(\hbar\omega_P)(\hbar\omega_S)^3(\hbar\omega_P-\hbar\omega_S)^3|\chi^{(2)}|^2 \quad , \qquad (2)$$

where $\chi^{(2)}$ is the X-ray susceptibility of atoms, which can be determined by methods described in [23].

Now let us consider head-on collisions of laser optical photons with hydrogen-like RI with energy levels $(E_2,E_1)$ in RF. The transition energy is equal to $E_2-E_1=\hbar\omega_{21}$. Suppose that now the resonance condition $2\gamma\hbar\omega_L=\hbar\omega_{21}$ takes place. Since the total cross sections of processes are invariants, the above described fluorescence radiation as well as the production of X-ray entangled photons will take place with the same cross sections both in LF and RF of the ions. However, the energies of the produced photons in LF and RF of RIs will be different. It is clear that in the LF, due to Doppler effect, the produced photons will have energies increased up to $4\gamma^2\omega_L$ and they will be emitted in a narrow cone with small opening angle of the order of $\sim 1/\gamma$. Let us remember that the total cross sections and the number of the produced photons are the same in LF and RF.

We will now do some numerical estimates. Let us suppose that, as in [18], the hydrogen-like ions are oxygen ions OVII having allowed transitions $1s^2{}^1S_0\to 1s2p\,^1P_1$ with parameters: $\hbar\omega_{21}=571.3\text{eV}, \Gamma_{21}=0.01364\text{eV}$. If the counter propagating laser beam is of argon laser (second case: Nd:YAG laser) with $\hbar\omega_L\approx 2.4$ eV ($\hbar\omega_L\approx 1.06$ eV), then resonance occurs at $\gamma=119$ ($\gamma=268.5$), and the produced single fluorescence photons will have energies up to $\hbar\omega_{Fl}\approx 4\gamma^2\omega_L\approx 136\text{keV}$ ($\hbar\omega_{Fl}\approx 4\gamma^2\omega_L\approx 307$ keV), while the energies of the entangled photons in the most probable case will be equal to $\hbar\omega_S\approx\hbar\omega_i\approx 68$ keV ($\hbar\omega_S\approx\hbar\omega_i\approx 153.5$ keV). Such ion beams could be obtained at RHIC and SPS by the method proposed in [20]. In the near future, RI will be obtained at LHC with higher Lorentz factors up to $\gamma\sim 2700$, which will allow to use atomic as well as nuclear levels for production of fluorescence and entangled photons with energies much higher than $10^5$ eV.

The energy spread $\Delta\gamma/\gamma \sim 10^{-3}$ of the RIs is usually is much larger than the energy spread $\Delta\omega_L/\omega_L$ of the laser photons. In [13, 18] it is shown that in this case the production of single fluorescence photons takes place with effective cross section $\sigma_{eff}^F = \sigma_{res}^F (\Gamma_{21}/\hbar\omega_{21})/(\Delta\gamma/\gamma)$, where $\sigma_{res}^F$ is the resonance fluorescent cross section (for the transition under consideration equal to $\sigma_{res}^F = 5.37 \cdot 10^{-16}$ cm² and $\sigma_{eff}^F = 2.9 \cdot 10^{-17}$ cm²; compare these numbers to the total cross section of Thomson scattering $6.65 \cdot 10^{-26}$ cm²). To obtain the cross section of entangled photon production with the help of (2), one needs to have the value of $\chi^{(2)}$.

Postponing the development of a correct theory and calculations, let us estimate $\chi^{(2)}$ using the results [22]: only the contribution of dipole transitions shall be accounted for. We take $\omega_S \approx \omega_i \approx \omega_P/2$. Then

$$\chi^{(2)} = \frac{\alpha^2 r_e}{2}\left(\frac{\hbar c}{\hbar \omega_S}\right)^3 \frac{e}{\Gamma_{21}}, \tag{3}$$

where $\alpha = 1/137$ is the fine structure constant, $e$ and $r_e$ are the electron charge and classic radius. For oxigen ions OVII transitions, $\chi^{(2)} = 5.66 \cdot 10^{-35}$ cm⁵(stC)⁻¹, where stC is the Gaussian unit of the electric charge. Using these values, we obtain the entangled photon resonance and effective production cross sections:

$$\sigma_{res}^{Ent}(0.57\text{keV} \to 0.285 \text{ keV} + 0.285 \text{ keV}) \approx 1.1 \ 10^{-21} \text{ cm}^2. \tag{4}$$

$$\sigma_{eff}^{Ent} \approx 2.63 \ 10^{-23} \text{ cm}^2. \tag{5}$$

As it follows, the production cross sections for X-ray entangled photons are $\sim 10^5$-$10^6$ times less than those for "background" fluorescence photons.

Now let us estimate the production rates, i.e. the number of fluorescence and X-ray entangled photon pair's produced per second in collisions of argon laser pulses with RHIC's RI bunches assuming moderate intensities. The number of produced photons per second is equal to $N_i = L\sigma_i$, where $\sigma_i$ is the cross section of the relevant process, luminosity $L$ is roughly equal to $L = f \cdot N_{RI} \cdot N_L/A$, $f$ is the frequency of the laser pulses synchronized with the bunches of RHIC, $N_{RI}$ and $N_L$ are the numbers of particles, $A$ is the geometrical cross section of RI bunches and laser pulses, respectively. Taking $N_{RI} \approx 10^9$, $A \approx \pi R_{RI}^2 \approx 7 \cdot 10^{-4}$ cm² (according to [23], the RHIC bunch radius is $R_{RI} \approx 0.015$ cm), and an argon laser operating with frequency $f = 1$ s⁻¹ and pulse energy ~1 mJ, and thus providing $N_L \approx 2.6 \cdot 10^{15}$, one obtains $L \approx 4 \cdot 10^{27} s^{-1} cm^{-2}$ and photon production rates $N^{Ent} \approx 9.5 \cdot 10^5$ s$^{-1}$ and $N^F \approx 1.1 \cdot 10^{11}$ s$^{-1}$. Focusing of the argon laser's beam to the size of RHIC's beam is easily achievable.

As it has been mentioned above, such a large number of ~68 keV entangled photon pairs is emitted in the forward direction of RI momentum in a cone with opening angle $\theta \approx 1/\gamma = 1/119 =$ 8.4 mrad. One can detect these entangled photons among the huge "background" of ~136 keV fluorescence photons using time and energy discrimination with the help of two X-ray detectors operating in coincidence and having medium energy resolution. At lower laser photon intensity one can use detector similar to one described in [24], which is able to measure also the polarization of the entangled photons.

Let us note that the above estimated rate of production of X-ray entangled photons, which is much higher than the obtained rates, can be further increased, as follows. It has been shown in [13,16,18] that for the RI OVII collisions with Nd:YAG laser's photons, when the laser photon density is higher than the available $n_L \geq 1.5\ 10^{15}$ cm$^{-3}$ (or intensity $I_L \geq 7.5\ 10^6$ Wcm$^{-2}$), stimulated fluorescence, and, therefore, entangled photon radiation will take place. It is also well known that X-ray entangled photon production is accompanied by emission of soft photons (see, for instance, [22]).

In conclusion, we have found a mechanism and made numerical estimates of entangled photons' production in ion – photon colliding beams. At available ion acceleration facilities and with available lasers, the energies of photons lie in hundred keV range, with number of photons produced being far larger than current numbers (rates).

The authors thank M.G. Aginian, A.T. Margarian, A. Mooradian and A. Tishchenko for discussions.